\documentclass{article}
\usepackage{amsmath, amsfonts, amssymb, amsthm}
\begin{document}

\begin{center}
{\large\bf
Analytic formulae for T violation in neutrino oscillations}
\end{center}
\vspace*{.6cm}

\begin{center}
\large{\sc Osamu Yasuda}
\end{center}
\vspace*{0cm}
{\it
\begin{center}
Department of Physics, Tokyo Metropolitan University,\\
Hachioji, Tokyo 192-0397, Japan
\end{center}}

\vspace*{0.5cm}

{\Large \bf
\begin{center} Abstract \end{center}  }
Recently, a concept known as $\mu$TRISTAN, which involves the
acceleration of $\mu^+$, has been proposed.  This initiative has led to
considerations of a new design for a neutrino factory.  Additionally,
leveraging the polarization of $\mu^+$, measurements of T violation in
neutrino oscillations are also being explored.  In this paper, we
present analytical expressions for T violation in neutrino
oscillations within the framework of standard three flavor neutrino
oscillations, a scenario involving nonstandard interactions, and a case
of unitarity violation.  We point out that examining the energy spectrum
of T violation may be useful for probing new physics effects.
\vspace*{0.5cm}

\section{Introduction}
Results from various neutrino oscillation experiments have nearly
determined the three mixing angles and the absolute values of the mass
squared differences in the standard three flavor mixing scenario
within the lepton sector\,\cite{ParticleDataGroup:2022pth}.  The
remaining undetermined parameters, such as the mass ordering, the
octant of the atmospheric neutrino oscillation mixing angle, and the
CP phase, are expected to be resolved by the high-intensity neutrino
long-baseline experiments currently under construction, such as T2HK
and DUNE.  Once the CP phase is established, the standard three flavor
lepton mixing scheme will be solidified, {achieving the final goal of
studies on standard three-flavor neutrino oscillations.}  To explore
physics beyond this framework using neutrino oscillations, experiments
in previously unexplored channels will be necessary.

Recently, a concept known as $\mu$TRISTAN\,\cite{Hamada:2022mua} has
been proposed, which involves creating a low-emittance $\mu^+$ beam
using ultra-cold muon technology and accelerating it to energies
suitable for a $\mu^+$ collider.  {The expected number of muons at
$\mu$TRISTAN is on the order of $10^{13}$ to $10^{14}$ muons per
second.}  This proposal has reignited
interest\,\cite{sugama} in the neutrino factory
concept\,\cite{Geer:1997iz,ISSPhysicsWorkingGroup:2007lul},
which could be developed en route to
achieving a muon collider.  At such a neutrino factory, the decay of
$\mu^+$ in the storage ring would produce $\overline{\nu}_\mu$ and $\nu_e$.
Ref.\,\cite{Ota:2001cz} explored the potential to polarize the $\mu$
beam to reduce the flux of $\nu_\mu$ or $\overline{\nu}_\mu$, thereby
enabling the measurement of $\nu_e\to\nu_\mu$ transitions.  If this can
be achieved, it would allow for the measurement of T violation in
neutrino oscillations, i.e., the difference between the oscillation
probabilities $P(\nu_\mu\to\nu_e)$ and $P(\nu_e\to\nu_\mu)$.

T violation in neutrino oscillations has been discussed by many
researchers in the
past\,\cite{Cabibbo:1977nk,Krastev:1988yu,Toshev:1991ku,Naumov:1991ju,Arafune:1996bt,Barger:1999hi,Koike:1999hf,Koike:1999tb,Harrison:1999df,Yokomakura:2000sv,Parke:2000hu,Koike:2000jf,Miura:2001pia,Akhmedov:2001kd,Miura:2001pi,Ota:2001cz,Guo:2001yt,Bueno:2001jd,Minakata:2002qi,Xing:2013uxa,Petcov:2018zka,Schwetz:2021cuj,Schwetz:2021thj}.
T violation has attracted significant attention primarily because its
structure is simpler than that of CP violation, which compares
$P(\nu_\mu\to\nu_e)$ with $P(\overline{\nu}_\mu\to\overline{\nu}_e)$
and involves complications due to the presence of the matter
effect.\footnote{To justify discussions of T violation on an equal footing
with CP violation, CPT symmetry is necessary in the neutrino sector.
Refs.\,\cite{deGouvea:2017yvn,Barenboim:2017ewj,Tortola:2020ncu,Ngoc:2022uhg,Barenboim:2023krl} studied CPT symmetry in the neutrino sector and
concluded that there are strong constraints on CPT violation.}
In this paper, we derive the analytical forms of T violation in three
scenarios: the standard three flavor scheme, a scenario with
nonstandard interactions, and a case of unitarity violation.  We also
briefly comment on the feasibility of probing new physics effects by
examining the energy dependence of T violation.

In Sect. \ref{generalities}, we review the formalism by Kimura,
Takamura, and Yokomakura\,\cite{Kimura:2002hb,Kimura:2002wd} to derive
analytical formulas for the oscillation probabilities.  In
Sect. \ref{t-violation}, we derive the analytic forms for T violation
in the three cases: the standard three flavor mixing framework, a
scenario involving flavor-dependent nonstandard interactions, and a
case with unitarity violation.  In Sect. \ref{conclusions}, we
summarize our conclusions.

\section{Analytical formula for oscillation probabilities \label{generalities}}

It has been known\,\cite{Grimus:1993fz} (See also
earlier works\,\cite{Halprin:1986pn,Mannheim:1987ef,Sawyer:1990tw}.)
that after eliminating the negative energy states
by a Tani-Foldy-Wouthusen-type transformation, the
Dirac equation for neutrinos propagating in matter is reduced to the
familiar form:
\begin{eqnarray}
&{\ }&\hspace*{-45mm}
  i\frac{d\Psi}{dt}=
\left(U{\cal E}U^{-1}
+{\cal A}
\right)\Psi,
\label{sch1}
\end{eqnarray}
where $U$ is the PMNS matrix, 
\begin{eqnarray}
  &{\ }&\hspace*{-60mm}
  \Psi\equiv\left(\begin{array}{c}
    \nu_e\\
    \nu_\mu\\
    \nu_\tau
    \end{array}\right)
\nonumber
\end{eqnarray}
is the flavor eigenstate,
\begin{eqnarray}
&{\ }&\hspace*{-50mm}
{\cal E}\equiv{\mbox{\rm diag}}\left(E_1,E_2,E_3\right)
\label{e}
\end{eqnarray}
is the diagonal matrix of the energy eigenvalue
$E_j\equiv\sqrt{m_j^2 +\vec{p}^{\,2}}~(j=1,2,3)$ of each
mass eigenstate with momentum $\vec{p}$,
and the matrix
\begin{eqnarray}
&{\ }&\hspace*{-15mm}
\displaystyle{\cal A}\equiv
\sqrt{2}\,G_F\,\left\{\mbox{\rm diag}
\left(N_e-\frac{N_n}{2},-\frac{N_n}{2},-\frac{N_n}{2}
\right)\right\}\,.
\nonumber
\end{eqnarray}
stands for the matter effect, which is
characterized by the Fermi coupling
constant $G_F$, the electron density
$N_e$ and the neutron density $N_n$.
Throughout this paper we assume for simplicity that the density of
matter is constant.
The $3\times3$ matrix on the
right hand side of Eq. (\ref{sch1}) is hermitian and can be formally
diagonalized by a unitary matrix $\widetilde{U}$ as:
\begin{eqnarray}
&{\ }&\hspace*{-45mm}
U{\cal E}U^{-1}+{\cal A}
=\widetilde{U}\widetilde{{\cal E}}\widetilde{U}^{-1},
\label{sch3}
\end{eqnarray}
where
\begin{eqnarray}
&{\ }&\hspace*{-50mm}
\widetilde{{\cal E}}\equiv{\mbox{\rm diag}}\left(
\widetilde{E}_1,\widetilde{E}_2,\widetilde{E}_3\right)
\nonumber
\end{eqnarray}
is a diagonal matrix with the energy eigenvalue
$\widetilde{E}_j$ in the presence of the matter effect.
Eq. (\ref{sch1}) can be easily solved, resulting
the flavor eigenstate at the distance $L$:
\begin{eqnarray}
&{\ }&\hspace*{-35mm}
\Psi(L)=\widetilde{U}\exp\left(-i\widetilde{{\cal E}}L\right)\widetilde{U}^{-1}\Psi(0).
\label{sol1}
\end{eqnarray}
Thus we have the probability amplitude
$A(\nu_\beta\to\nu_\alpha)$ of the flavor transition $\nu_\beta\to\nu_\alpha$:
\begin{eqnarray}
  &{\ }&\hspace*{-35mm}
  A(\nu_\beta\to\nu_\alpha)
  =\left[\widetilde{U}
  \exp\left(-i\widetilde{{\cal E}}L\right)
  \widetilde{U}^{-1}\right]_{\alpha\beta}.
\label{sol2}
\end{eqnarray}
From Eq. (\ref{sol2}) we observe that
the shift ${\cal E}\to{\cal E}-{\bf 1}E_1$,
where ${\bf 1}$ stands for the $3\times 3$ identity matrix,
changes only the overall phase of the probability amplitude
$A(\nu_\beta\to\nu_\alpha)$, and this phase
does not affect the value of the probability 
$P(\nu_\beta\to\nu_\alpha) = |A(\nu_\beta\to\nu_\alpha)|^2$
of the flavor transition $\nu_\beta\to\nu_\alpha$.
In the following discussions,
therefore, for simplicity, we define the
diagonal energy matrix ${\cal E}$ and
the potential one ${\cal A}$ as follows:
\begin{eqnarray}
&{\ }&\hspace*{-5mm}
  {\cal E}\equiv{\mbox{\rm diag}}\left(E_1,E_2,E_3\right)
  -E_1\,{\bf 1}
\nonumber\\
&{\ }&\hspace*{-2mm}
  ={\mbox{\rm diag}}\left(0,\Delta E_{21},\Delta E_{31}\right)
\label{em}\\
&{\ }&\hspace*{-5mm}
{\cal A}\equiv\sqrt{2}\,G_F\,\left\{\mbox{\rm diag}
\displaystyle
\left(N_e-\frac{N_n}{2},-\frac{N_n}{2},-\frac{N_n}{2}
\right)+\left(\frac{N_n}{2}\right)\,{\bf 1}\right\}
\nonumber\\
&{\ }&\hspace*{-2mm}
  ={\mbox{\rm diag}}\left(A,0,0\right)
\label{am}\\
&{\ }&\hspace*{-32mm}
\mbox{\rm with}
\nonumber\\
&{\ }&\hspace*{-5mm}
\Delta E_{jk}\equiv E_j - E_k\simeq \frac{m_j^2-m_k^2}
{2|\vec{\relax{\kern .1em p}}|}\equiv
\frac{\Delta m_{jk}^2}{2|\vec{\relax{\kern .1em p}}|}
\equiv\frac{\Delta m_{jk}^2}{2E}
\nonumber\\
&{\ }&\hspace*{-5mm}
A\equiv \sqrt{2}\,G_F\,N_e\,.
\label{a}
\end{eqnarray}
Thus the appearance oscillation probability
$P(\nu_\beta\to\nu_\alpha)~(\alpha\ne\beta)$ is
given by
\begin{eqnarray}
&{\ }&\hspace*{-11mm}
  P(\nu_\beta\to\nu_\alpha)
\nonumber\\
&{\ }&\hspace*{-16mm}
  =
\left|\left[\widetilde{U}\exp\left(-i\widetilde{{\cal E}}L\right)
    \widetilde{U}^{-1}
    \right]_{\alpha\beta}\right|^2
\nonumber\\
&{\ }&\hspace*{-16mm}
  =\left|
\sum_{j=1}^3 \widetilde X_j^{\alpha\beta}e^{-i\widetilde E_j L}
\right|^2
\nonumber\\
&{\ }&\hspace*{-16mm}
= \left|e^{-i\widetilde E_1 L}
\sum_{j=1}^3 \widetilde X_j^{\alpha\beta}e^{-i\Delta \widetilde E_{j1} L}
\right|^2
\nonumber\\
&{\ }&\hspace*{-16mm}
= \left|
\sum_{j=1}^3 \widetilde X_j^{\alpha\beta}\left(e^{-i\Delta \widetilde E_{j1} L}-1\right)
\right|^2
\label{probv0}
\\
&{\ }&\hspace*{-16mm}
= \left|
(-2i)\sum_{j=2}^3
e^{-i\Delta \widetilde E_{j1} L/2}
\widetilde X_j^{\alpha\beta}
\sin\left(\frac{\Delta \widetilde E_{j1} L}{2}\right)
\right|^2
\nonumber\\
&{\ }&\hspace*{-16mm}
= 4\left|
e^{-i\Delta \widetilde E_{31} L/2}
\widetilde X_3^{\alpha\beta}
\sin\left(\frac{\Delta \widetilde E_{31} L}{2}\right)
+e^{-i\Delta \widetilde E_{21} L/2}
\widetilde X_2^{\alpha\beta}
\sin\left(\frac{\Delta \widetilde E_{21} L}{2}\right)
\right|^2
\nonumber\\
&{\ }&\hspace*{-16mm}
=4 \left|
\widetilde X_3^{\alpha\beta}
\sin\left(\frac{\Delta \widetilde E_{31} L}{2}\right)
+e^{i\Delta \widetilde E_{32} L/2}
\widetilde X_2^{\alpha\beta}
\sin\left(\frac{\Delta \widetilde E_{21} L}{2}\right)
\right|^2
\label{probv}
\end{eqnarray}
where
\begin{eqnarray}
&{\ }&\hspace*{-75mm}
  \widetilde{X}^{\alpha\beta}_j
  \equiv\widetilde{U}_{\alpha j}\widetilde{U}^\ast_{\beta j}
\nonumber\\
&{\ }&\hspace*{-75mm}
\Delta \widetilde{E}_{jk}
\equiv\widetilde{E}_j-\widetilde{E}_k
\nonumber
\end{eqnarray}
have been defined,
\begin{eqnarray}
&{\ }&\hspace*{-55mm}
\sum_{j=1}^3 \widetilde X_j^{\alpha\beta}=\delta_{\alpha\beta}=0
~~\mbox{\rm for}~\alpha\ne\beta
\label{unitarity}
\end{eqnarray}
was subtracted in Eq. (\ref{probv0})
and throughout this paper the indices $\alpha, \beta = (e, \mu, \tau)$
and $j, k = (1, 2, 3)$ stand for those of the flavor and mass eigenstates,
respectively.
Once we know the eigenvalues $\widetilde{E}_j$ and the
quantity $\widetilde{X}^{\alpha\beta}_j$, the oscillation
probability can be expressed analytically.\footnote{
In the case of three neutrino flavors in matter, the energy
eigenvalues, $\widetilde{E}_j$, can in principle be analytically
determined using the cubic equation root
formula\,\cite{Barger:1980tf}.  However, the analytic expression for
$\widetilde{E}_j$ involving the inverse cosine function is not
practically useful.  Therefore, below we will calculate
$\widetilde{E}_j$ using perturbation theory with small parameters, such
as $\Delta E_{21}/\Delta E_{31} = \Delta m_{21}^2/\Delta m_{31}^2 \approx
1/30$ and those relevant to nonstandard scenarios.
}
So the only non-trivial problem in the standard case
is to obtain the expression for $\widetilde{X}^{\alpha\beta}_j$,
and this was done by Kimura, Takamura and
Yokomakura\,\cite{Kimura:2002hb,Kimura:2002wd}.
Their arguments are based on the trivial identities.
From the unitarity condition of the matrix $\widetilde{U}$,
we have
\begin{eqnarray}
&{\ }&\hspace*{-40mm}
\delta_{\alpha\beta}=\left[\widetilde{U}\widetilde{U}^{-1}\right]_{\alpha\beta}
=\sum_j\widetilde{U}_{\alpha j}\widetilde{U}^\ast_{\beta j}
=\sum_j\widetilde{X}^{\alpha\beta}_j.
\label{const1}
\end{eqnarray}
Next we take the $(\alpha,\beta)$ component of the both hand sides
in Eq. (\ref{sch3}):
\begin{eqnarray}
&{\ }&\hspace*{-15mm}
\left[U{\cal E}U^{-1}+{\cal A}\right]_{\alpha\beta}
=\left[\widetilde{U}\widetilde{{\cal E}}\widetilde{U}^{-1}\right]_{\alpha\beta}
=\sum_j\widetilde{U}_{\alpha j}\widetilde{E}_j\widetilde{U}^\ast_{\beta j}
=\sum_j\widetilde{E}_j\widetilde{X}^{\alpha\beta}_j
\label{const2}
\end{eqnarray}
Furthermore, we take the $(\alpha,\beta)$ component of
the square of Eq. (\ref{sch3}):
\begin{eqnarray}
&{\ }&\hspace*{-5mm}
\left[\left(U{\cal E}U^{-1}+{\cal A}\right)^2\right]_{\alpha\beta}
=\left[\widetilde{U}\widetilde{{\cal E}}^2\widetilde{U}^{-1}\right]_{\alpha\beta}
=\sum_j\widetilde{U}_{\alpha j}\widetilde{E}^2_j\widetilde{U}^\ast_{\beta j}
=\sum_j\widetilde{E}^2_j\widetilde{X}^{\alpha\beta}_j
\label{const3}
\end{eqnarray}
Putting Eqs. (\ref{const1})--(\ref{const3}) together,
we have
\begin{eqnarray}
&{\ }&\hspace*{-40mm}
\left(\begin{array}{ccc}
1&1&1\cr
\widetilde{E}_1&\widetilde{E}_2&\widetilde{E}_3\cr
\widetilde{E}^2_1&\widetilde{E}^2_2&\widetilde{E}^2_3
\end{array}\right)
\left(\begin{array}{c}
\widetilde{X}^{\alpha\beta}_1\cr
\widetilde{X}^{\alpha\beta}_2\cr
\widetilde{X}^{\alpha\beta}_3
\end{array}\right)
=\left(\begin{array}{r}
Y_1^{\alpha\beta}\cr
Y_2^{\alpha\beta}\cr
Y_3^{\alpha\beta}
  \end{array}\right)
\label{verdermonde}\\
&{\ }&\hspace*{-58mm}
\mbox{\rm with}
\nonumber\\
  &{\ }&\hspace*{-40mm}
Y_j^{\alpha\beta}
\equiv
\left[\left(U{\cal E}U^{-1}+{\cal A}\right)^{j-1}\right]_{\alpha\beta}
\quad\mbox{\rm for}~~j=1,2,3\,,
\nonumber
\end{eqnarray}
which can be easily solved by inverting the
Vandermonde matrix:
\begin{eqnarray}
\left(\begin{array}{c}
\widetilde{X}^{\alpha\beta}_1\cr\cr
\widetilde{X}^{\alpha\beta}_2\cr\cr
\widetilde{X}^{\alpha\beta}_3
\end{array}\right)
=\left(\begin{array}{ccc}
\displaystyle
\frac{{\ }1}{\Delta \widetilde{E}_{21} \Delta \widetilde{E}_{31}}
(\widetilde{E}_2\widetilde{E}_3, & -(\widetilde{E}_2+\widetilde{E}_3),&
1)\cr
\displaystyle
\frac{-1}{\Delta \widetilde{E}_{21} \Delta \widetilde{E}_{32}}
(\widetilde{E}_3\widetilde{E}_1, & -(\widetilde{E}_3+\widetilde{E}_1),&
1)\cr
\displaystyle
\frac{{\ }1}{\Delta \widetilde{E}_{31} \Delta \widetilde{E}_{32}}
(\widetilde{E}_1\widetilde{E}_2, & -(\widetilde{E}_1+\widetilde{E}_2),&
1)\cr
\end{array}\right)
\left(\begin{array}{r}
Y_1^{\alpha\beta}\cr\cr
Y_2^{\alpha\beta}\cr\cr
Y_3^{\alpha\beta}
  \end{array}\right)\,.
\label{solx}
\end{eqnarray}

\section{Analytic form of T violation \label{t-violation}}
In this section, we derive the analytic form
of T violation in the cases with and without unitarity,
using the formalism described in Sect. \ref{generalities}.

\subsection{The three flavor case with unitarity \label{unitary}}
First let us discuss the case where time evolution is unitary.
From Eq. (\ref{probv}), we have
\begin{eqnarray}
  &{\ }&\hspace*{-25mm}
  P(\nu_\mu\to\nu_e)-P(\nu_e\to\nu_\mu)
\nonumber\\
&{\ }&\hspace*{-30mm}
=4\sin\left(\frac{\Delta \widetilde E_{31} L}{2}\right)
\sin\left(\frac{\Delta \widetilde E_{21} L}{2}\right)
\nonumber\\
&{\ }&\hspace*{-27mm}
\times \left[
  e^{i\Delta \widetilde E_{32} L/2}\widetilde X_3^{e\mu\ast}\widetilde X_2^{e\mu}
+  e^{-i\Delta \widetilde E_{32} L/2}\widetilde X_3^{e\mu}\widetilde X_2^{e\mu\ast}\right.
\nonumber\\
&{\ }&\hspace*{-27mm}
\left.
- e^{i\Delta \widetilde E_{32} L/2}\widetilde X_3^{e\mu}\widetilde X_2^{e\mu\ast}
- e^{-i\Delta \widetilde E_{32} L/2}\widetilde X_3^{e\mu\ast}\widetilde X_2^{e\mu}
\right]
\nonumber\\
&{\ }&\hspace*{-30mm}
=16\,
\mbox{\rm Im}\,\left[\widetilde X_2^{e\mu}
  \widetilde X_3^{e\mu\ast}\right]\,
\sin\left(\frac{\Delta \widetilde E_{32} L}{2}\right)
\sin\left(\frac{\Delta \widetilde E_{31} L}{2}\right)
\sin\left(\frac{\Delta \widetilde E_{21} L}{2}\right)
\,,
\label{tv1}
\end{eqnarray}
Furthermore, from Eq. (\ref{solx}),
the factor $\mbox{\rm Im}\,\left[\widetilde X_2^{e\mu}
  \widetilde X_3^{e\mu\ast}\right]$ in Eq. (\ref{tv1})
can be rewritten as
\begin{eqnarray}
&{\ }&\hspace*{-20mm}
\mbox{\rm Im}\,\left[\widetilde X_2^{e\mu}
  \widetilde X_3^{e\mu\ast}\right]
=\displaystyle
\frac{-1}{\Delta \widetilde{E}_{21} \Delta \widetilde{E}_{32}}
\frac{{\ }1}{\Delta \widetilde{E}_{31} \Delta \widetilde{E}_{32}}
\nonumber\\
&{\ }&\hspace*{5mm}
\times\,\mbox{\rm Im}\,\left[
  \{Y_3^{e\mu}-(\widetilde{E}_3+\widetilde{E}_1)Y_2^{e\mu}\}
  \{Y_3^{e\mu\ast}-(\widetilde{E}_1+\widetilde{E}_2)Y_2^{e\mu\ast}\}\right]
\nonumber\\
&{\ }&\hspace*{2mm}
=\displaystyle
\frac{1}{\Delta \widetilde{E}_{21} \Delta \widetilde{E}_{31}
  \Delta \widetilde{E}_{32}}
\,\mbox{\rm Im}\,\left[Y_2^{e\mu}Y_3^{e\mu\ast}\right]
\label{tv2}
\end{eqnarray}
Eqs. (\ref{tv1})
and (\ref{tv2}) are applicable for a generic case,
as long as unitarity relation (\ref{unitarity}) holds.

\subsubsection{The standard three flavor case \label{standard}}
In the standard three flavor case,
$Y_{j+1}^{\alpha\beta}
\equiv[(U{\cal E}U^{-1}+{\cal A})^j]_{\alpha\beta}$ $(j=1,2)$
can be expressed as follows:
\begin{eqnarray}
  &{\ }&\hspace*{-30mm}
  Y_2^{\alpha\beta}\equiv
  \left[U{\cal E}U^{-1}+{\cal A}\right]_{\alpha\beta}
\nonumber\\
&{\ }&\hspace*{-23mm}
  =
\sum_{j=2}^3 \Delta E_{j1} X^{\alpha\beta}_j+ A\,\delta_{{\alpha}e}\delta_{{\beta}e}
\label{stdy2}\\
&{\ }&\hspace*{-30mm}
Y_3^{\alpha\beta}\equiv
\left[\left(U{\cal E}U^{-1}+{\cal A}\right)^2\right]_{\alpha\beta}
\nonumber\\
&{\ }&\hspace*{-23mm}
=
\sum_{j=2}^3 (\Delta E_{j1})^2 X^{\alpha\beta}_j
+A \sum_{j=2}^3 \Delta E_{j1} \left(\delta_{{\alpha}e}X^{e\beta}_j
+\delta_{{\beta}e}X^{{\alpha}e}_j\right)
+ A^2\,\delta_{{\alpha}e}\delta_{{\beta}e}\,,
\label{stdy3}
\end{eqnarray}
where we have also defined the quantity in vacuum:
\begin{eqnarray}
&{\ }&\hspace*{-95mm}
  X^{\alpha\beta}_j
  \equiv U_{\alpha j} U^\ast_{\beta j}\,.
\label{xab}
\end{eqnarray}
From Eqs. (\ref{stdy2}) and (\ref{stdy3}),
the factor 
$\mbox{\rm Im}\,\left[Y_2^{e\mu}Y_3^{e\mu\ast}\right]$
in Eq. (\ref{tv2}) can be rewritten as
\begin{eqnarray}
  &{\ }&\hspace*{-32mm}
  \mbox{\rm Im}\,\left[Y_2^{e\mu}Y_3^{e\mu\ast}\right]
\nonumber\\
&{\ }&\hspace*{-35mm}
=\mbox{\rm Im}\,\left[
  \left(\Delta E_{21}X_2^{e\mu}+\Delta E_{31}X_3^{e\mu}\right)\right.
\nonumber\\
&{\ }&\hspace*{-26mm}
  \left.\times\left\{\Delta E_{21}\left(\Delta E_{21}+A\right)X_2^{e\mu\ast}
  +\Delta E_{31}\left(\Delta E_{31}+A\right)X_3^{e\mu\ast}\right\}
  \right]
\nonumber\\
&{\ }&\hspace*{-35mm}
=\mbox{\rm Im}\,\left[X_2^{e\mu}X_3^{e\mu\ast}\right]
\Delta E_{21}\Delta E_{31}\Delta E_{32}
\label{stdtv1}
\end{eqnarray}
$\mbox{\rm Im}\,\left[X_2^{e\mu}X_3^{e\mu\ast}\right]$
in Eq. (\ref{stdtv1})
is the Jarlskog factor\,\cite{Jarlskog:1985ht} for the lepton sector, and is
given in the standard parametrization\,\cite{ParticleDataGroup:2022pth}
with the three mixing angles $\theta_{jk}~(j,k=1,2,3)$ and
the Dirac CP phase $\delta$ by
\begin{eqnarray}
&{\ }&\hspace*{-50mm}
J\equiv\,\mbox{\rm Im}\,\left[X_2^{e\mu}X_3^{e\mu\ast}\right]
\nonumber\\
&{\ }&\hspace*{-47mm}
=-\frac{1}{8}\sin\delta\cos\theta_{13}
\sin 2\theta_{12}\sin 2\theta_{13}\sin 2\theta_{23}\,.
\nonumber
\end{eqnarray}
Hence we obtain
\begin{eqnarray}
  &{\ }&\hspace*{-27mm}
  P(\nu_\mu\to\nu_e)-P(\nu_e\to\nu_\mu)
  \nonumber\\
&{\ }&\hspace*{-30mm}
  = 16\,J\,\frac{\Delta E_{21}\Delta E_{31}\Delta E_{32}}
  {\Delta \widetilde{E}_{21} \Delta \widetilde{E}_{31}
  \Delta \widetilde{E}_{32}}\,
\sin\left(\frac{\Delta \widetilde E_{32} L}{2}\right)
\sin\left(\frac{\Delta \widetilde E_{31} L}{2}\right)
\sin\left(\frac{\Delta \widetilde E_{21} L}{2}\right)
\label{stdtv2}
\end{eqnarray}
Eq. (\ref{stdtv2}) is a well known formula\,\cite{Naumov:1991ju}
for the standard three flavor case.
It is remarkable that in the standard three flavor case,
T violation in matter{, when divided by the $\delta$-independent
factor $16\,\prod_{j>k}
[\sin(\Delta \widetilde{E}_{jk}L/2)\Delta E_{jk}/\Delta \widetilde{E}_{jk}]$,
coincides with} the Jarlskog factor in vacuum.  This
implies that the only source of T violation is the CP phase $\delta$
in the standard three flavor case, and it is the reason why T
violation is simpler than CP violation in neutrino oscillations.

\subsubsection{The case with non-standard interactions \label{nonstandard}}
As long as unitarity in the three flavor framework is maintained,
Eq. (\ref{tv2}) holds.  In this subsection, let us consider the
scenario with flavor-dependent nonstandard
interactions\,\cite{Guzzo:1991hi,Roulet:1991sm} during neutrino
propagation.  This scenario has garnered significant attention due to
its potential implications for phenomenology.
In this case the mass matrix is given by
\begin{eqnarray}
&{\ }&\hspace*{-60mm}
U{\cal E}U^{-1}+{\cal A}
+{\cal A}_{NP}
\label{matrixnp}
\end{eqnarray}
with
\begin{eqnarray}
&{\ }&\hspace*{-45mm}
{\cal A}_{NP}\equiv A
\left(
\begin{array}{ccc}
\epsilon_{ee} & \epsilon_{e\mu} & \epsilon_{e\tau}\\
 \epsilon_{e\mu}^\ast & \epsilon_{\mu\mu} & \epsilon_{\mu\tau}\\
 \epsilon_{e\tau}^\ast & \epsilon_{\mu\tau}^\ast & \epsilon_{\tau\tau}
\end{array}
\right)\,,
\nonumber
\end{eqnarray}
where ${\cal A}$ and $A$ are given by Eqs. (\ref{am}) and (\ref{a}),
respectively.  The dimensionless quantities
$\epsilon_{\alpha\beta}$ stand for the ratio
of the nonstandard Fermi coupling constant
interaction to the standard one.
Since the matrix (\ref{matrixnp}) is hermitian,
time evolution is unitary and all the arguments
up to Eq. (\ref{tv2}) hold also in this case.
The oscillation probability
is given by Eqs. (\ref{probv}) and (\ref{solx}),
where the standard potential matrix ${\cal A}$
must be replaced by ${\cal A}+{\cal A}_{NP}$.

The extra complication compared to the standard case
is calculations of the eigenvalues $\widetilde{E}_j$
and the elements $[(U{\cal E}U^{-1}+{\cal A}+{\cal A}_{NP})^m]_{\alpha\beta}$
($m=1,2$).  Here we work with perturbation theory
with respect to the small parameters
$\Delta E_{21}/\Delta E_{31} = \Delta m_{21}^2/\Delta m_{31}^2\simeq 1/30$
and $\epsilon_{\alpha\beta}$, which we assume to be as small
as $\Delta E_{21}/\Delta E_{31}$.  Namely, throughout this paper
we assume
\begin{eqnarray}
  &{\ }&\hspace*{-40mm}
    |\Delta E_{31}| \sim A \gg |\Delta E_{21}|
\gtrsim A\,|\epsilon_{\alpha\beta}|\,.
\nonumber
\end{eqnarray}
and take into consideration to first order in
these small parameters.  Then, to first order in them, we get
\begin{eqnarray}
  &{\ }&\hspace*{-30mm}
  Y_2^{e\mu}\equiv
  \left[U{\cal E}U^{-1}+{\cal A}+{\cal A}_{NP}\right]_{e\mu}
\nonumber\\
&{\ }&\hspace*{-23mm}
=\Delta E_{31} X^{e\mu}_3 + \Delta E_{21} X^{e\mu}_2
+ A\,\epsilon_{e\mu}
\label{nsiy2}\\
&{\ }&\hspace*{-30mm}
Y_3^{e\mu}\equiv
\left[\left(U{\cal E}U^{-1}+{\cal A}+{\cal A}_{NP}\right)^2\right]_{e\mu}
\nonumber\\
&{\ }&\hspace*{-23mm}
\simeq\{(\Delta E_{31})^2+A\,\Delta E_{31}\} X^{e\mu}_3
+A\,\Delta E_{21} X^{e\mu}_2
\nonumber\\
&{\ }&\hspace*{-18mm}
+ A^2\,\epsilon_{e\mu}
+ \Delta E_{31} \left\{X_3, {\cal A}_{NP}\right\}_{e\mu}
\nonumber\\
&{\ }&\hspace*{-23mm}
=(\Delta E_{31}+A)\,Y_2^{e\mu} - \Delta E_{31}\Delta E_{21} X^{e\mu}_2
-A\,\Delta E_{31}\,\epsilon_{e\mu}
+ \Delta E_{31} \left\{X_3, {\cal A}_{NP}\right\}_{e\mu}\,,
\label{nsiy3}
\end{eqnarray}
where the curly bracket stands for an anticommutator
of matrices $P$ and $Q$:
$\{P,Q\}\equiv PQ+QP$, and
$X_3$ is a $3\times 3$ matrix defined by
\begin{eqnarray}
  &{\ }&\hspace*{-63mm}
  X_3\equiv U\,\mbox{\rm diag}\left(0,0,1\right)\,U^{-1}
\label{x3}\\
&{\ }&\hspace*{-63mm}
\left(X_3\right)_{\alpha\beta} = U_{\alpha 3} U_{\beta 3}^\ast
= X_3^{\alpha\beta}\,.
\nonumber
\end{eqnarray}
From this, the first term on the right hand side of
Eq. (\ref{nsiy3}) drops in the factor 
$\mbox{\rm Im}\,\left[Y_2^{e\mu}Y_3^{e\mu\ast}\right]$,
and we obtain:
\begin{eqnarray}
  &{\ }&\hspace*{-13mm}
  \mbox{\rm Im}\,\left[Y_2^{e\mu}Y_3^{e\mu\ast}\right]
  =-\mbox{\rm Im}\,\left[Y_2^{e\mu\ast}Y_3^{e\mu}\right]
\nonumber\\
&{\ }&\hspace*{-17mm}
\simeq \Delta E_{31}\,\mbox{\rm Im}\,\left[Y_2^{e\mu\ast}
  \left(\Delta E_{21} X^{e\mu}_2
  + A\,\epsilon_{e\mu}
  -\left\{X_3, {\cal A}_{NP}\right\}_{e\mu}
  \right)\right]
\nonumber\\
&{\ }&\hspace*{-17mm}
\simeq (\Delta E_{31})^2
\,\mbox{\rm Im}\,\left[X_3^{e\mu\ast}
  \left(\Delta E_{21} X^{e\mu}_2 + A\,\epsilon_{e\mu}
  -\left\{X_3, {\cal A}_{NP}\right\}_{e\mu}
  \right)\right]
\nonumber\\
&{\ }&\hspace*{-17mm}
= (\Delta E_{31})^2
\,\mbox{\rm Im}\,\left[X_3^{e\mu\ast}
  \left\{\Delta E_{21} X^{e\mu}_2 
+ A\,\left(X_3^{\tau\tau} \epsilon_{e\mu}
-X_3^{e\tau}\epsilon_{\tau\mu}
-X_3^{\tau\mu}\epsilon_{e\tau}\right)
  \right\}\right]\,,
\nonumber
\end{eqnarray}
where we have ignored terms of order
$O((\Delta E_{21})^2)$,
$O(A^2(\epsilon_{\alpha\beta})^2)$
and $O(A\Delta E_{21}\epsilon_{\alpha\beta})$.
Thus we finally get the form for T violation:
\begin{eqnarray}
&{\ }&\hspace*{-10mm}
  P(\nu_\mu\to\nu_e)-P(\nu_e\to\nu_\mu)
  \nonumber\\
&{\ }&\hspace*{-15mm}
  \simeq\displaystyle
\frac{16\,(\Delta E_{31})^2}{\Delta \widetilde{E}_{21} \Delta \widetilde{E}_{31}
  \Delta \widetilde{E}_{32}}
\sin\left(\frac{\Delta \widetilde E_{32} L}{2}\right)
\sin\left(\frac{\Delta \widetilde E_{31} L}{2}\right)
\sin\left(\frac{\Delta \widetilde E_{21} L}{2}\right)
\nonumber\\
&{\ }&\hspace*{-10mm}
\times 
\,\mbox{\rm Im}\,\left[X_3^{e\mu\ast}
  \left\{\Delta E_{21} X^{e\mu}_2 
  + A\,\left(X_3^{\tau\tau} \epsilon_{e\mu}
-\,X_3^{e\tau}\epsilon_{\tau\mu}
-\,X_3^{\tau\mu}\epsilon_{e\tau}\right)
  \right\}\right]
\label{nsitv}
\end{eqnarray}
Note that the form of the standard contribution $(\Delta
E_{31})^2\Delta E_{21},\mbox{\rm
Im}\,\left[X^{e\mu}_2X_3^{e\mu\ast}\right]$ in Eq. (\ref{nsitv})
differs slightly from that in Eq. (\ref{stdtv1}) because we are
neglecting terms of order $O((\Delta E_{21})^2)$.  The terms
proportional to $A$ in the parenthesis in Eq. (\ref{nsitv}) represent
the additional contributions to T violation due to nonstandard
interactions.  These additional contributions are constant with
respect to the neutrino energy $E$, and they exhibit a different
energy dependence from that of the standard one, $\Delta E_{21}
X^{e\mu}_2 = (\Delta m_{21}^2/2E)U_{e2} U_{\mu 2}^\ast$.  Therefore, if
the magnitude of the additional contributions from nonstandard
interactions is significant enough,\footnote{Constraints on the
parameters $\epsilon_{\alpha\beta}$ have been provided in
Refs.\,\cite{Davidson:2003ha,Biggio:2009nt,Coloma:2023ixt}.  Depending
on the sensitivity of each experiment, it may or may not be possible
to detect the signal or to improve the existing bounds on
$\epsilon_{\alpha\beta}$.  The aim of this paper is to derive the
analytic form of T violation; estimating experimental sensitivity is
beyond its scope.} then their effects are expected to be observable in the
energy spectrum of T violation.

\subsection{The three flavor case with unitarity violation \label{uv}}
The discussions in Subsect. \ref{unitary} are
based on the assumption that time evolution is unitary.
In Ref.\,\cite{Antusch:2006vwa},
the possibility to have a non-unitary leptonic mixing matrix
was pointed out.  In that case, the relation between
the mass eigenstate $\nu_j$ and the flavor eigenstate $\nu_\alpha$
is given by a nonunitary matrix $N$:
\begin{eqnarray}
  &{\ }&\hspace*{-63mm}
  \nu_\alpha = N_{\alpha j}\,\nu_j
\nonumber\\
&{\ }&\hspace*{-63mm}
N\equiv({\bf 1}+\eta)\, U 
\nonumber\\
&{\ }&\hspace*{-93mm}
\mbox{\rm with}
\nonumber\\
&{\ }&\hspace*{-63mm}
\eta^\dagger=\eta\,.
\label{eta}
\end{eqnarray}
In the so-called minimal unitarity violation, which was discussed
in Ref.\,\cite{Antusch:2006vwa}, the constraint on the deviation
matrix $\eta$ turned out to be strong.  Here we take
phenomenologically the form of the nonunitary matrix $N$
and assume that the elements of the deviation matrix $\eta$
is of order $\Delta E_{21}/\Delta E_{31}$ or smaller,
as in Subsect. \ref{nonstandard}, namely,
\begin{eqnarray}
  &{\ }&\hspace*{-40mm}
    |\Delta E_{31}| \sim A \gg |\Delta E_{21}|
\gtrsim A\,|\eta_{\alpha\beta}|\,.
\nonumber
\end{eqnarray}
It was argued in Ref.\,\cite{Fernandez-Martinez:2007iaa}
that time evolution in the case of nonunitary mixing matrix
can be discussed in terms of the mass eigenstate
\begin{eqnarray}
&{\ }&\hspace*{-65mm}
  \Psi_m\equiv\left(\begin{array}{c}
    \nu_1\\
    \nu_2\\
    \nu_3
    \end{array}\right)\,,
\nonumber
\end{eqnarray}
and its time evolution is described by
\begin{eqnarray}
&{\ }&\hspace*{-20mm}
  i\frac{d\Psi_m}{dt}=
\left\{{\cal E}
+N^T{\cal A}N^\ast
-A_n\left(N^TN^\ast-{\bf 1}\right)
\right\}\Psi_m\,,
\label{eomuv1}
\end{eqnarray}
where ${\cal E}$ and ${\cal A}$ are defined by
Eqs. (\ref{em}) and (\ref{am}),
\begin{eqnarray}
&{\ }&\hspace*{-65mm}
A_n\equiv\,\frac{1}{\sqrt{2}}G_F\,N_n
\nonumber
\end{eqnarray}
stands for the absolute value of
the contribution to the matter effect
from the neutral current interaction,
and the term $A_n\,{\bf 1}$ was added to simplify the
calculations without changing the absolute value
of the probability amplitude.
The $3\times3$ matrix on the
right hand side of Eq. (\ref{eomuv1}) can be
diagonalized with a unitary matrix $W$:
\begin{eqnarray}
&{\ }&\hspace*{-20mm}
  {\cal E}+N^T{\cal A}N^\ast
-A_n\left(N^TN^\ast-{\bf 1}\right)
  =W{\cal \widetilde{E}}W^{-1},
\label{uvm}
\end{eqnarray}
where
\begin{eqnarray}
&{\ }&\hspace*{-55mm}
\widetilde{{\cal E}}\equiv{\mbox{\rm diag}}\left(
\widetilde{E}_1,\widetilde{E}_2,\widetilde{E}_3\right)
\nonumber
\end{eqnarray}
is the energy eigenvalue matrix in matter
with unitarity violation.
The mass eigenstate at distance $L$ can be solved as
\begin{eqnarray}
&{\ }&\hspace*{-35mm}
\Psi_m(L)=W\exp(-i{\cal \widetilde{E}}L)W^{-1}\Psi_m(0).
\label{uvt}
\end{eqnarray}
In cases involving unitarity violation, due to the modified form of
the charged current interaction\,\cite{Antusch:2006vwa}, after
computing the probability amplitude from Eq. (\ref{uvt}), we must
multiply the probability amplitude by an additional factor of
$(NN^\dagger)_{\beta\beta}^{-1/2}$ for the production process and
$(NN^\dagger)_{\alpha\alpha}^{-1/2}$ for detection.
Defining the modified amplitude
\begin{eqnarray}
&{\ }&\hspace*{-20mm}
\hat{A}(\nu_\beta\to\nu_\alpha)
\equiv A(\nu_\beta\to\nu_\alpha)
(NN^\dagger)_{\alpha\alpha}^{1/2}
(NN^\dagger)_{\beta\beta}^{1/2}
\nonumber\\
&{\ }&\hspace*{-0mm}
=[N^\ast W\exp(-i{\cal \widetilde{E}}L)W^{-1}N^T]_{\alpha\beta}\,,
\nonumber
\end{eqnarray}
the modified probability
\begin{eqnarray}
&{\ }&\hspace*{-45mm}
\hat{P}(\nu_\alpha\to\nu_\beta)
\equiv|\hat{A}(\nu_\alpha\to\nu_\beta)|^2\,,
  \nonumber
\end{eqnarray}
and the quantity
\begin{eqnarray}
  &{\ }&\hspace*{-50mm}
  \widetilde{{\cal X}}^{\alpha\beta}_j\equiv
(N^{*}W)_{\alpha j}(NW^{*})_{\beta j}\,,
\nonumber
\end{eqnarray}
we have the following expression for
the appearance oscillation probability:
\begin{eqnarray}
&{\ }&\hspace*{-11mm}
  \hat{P}(\nu_\beta\to\nu_\alpha)
\nonumber\\
&{\ }&\hspace*{-16mm}
  =
  \left|\left[N^\ast W
    \exp\left(-i\widetilde{{\cal E}}L\right)
    W^{-1}N^T
    \right]_{\alpha\beta}\right|^2
\nonumber\\
&{\ }&\hspace*{-16mm}
  =\left|
\sum_{j=1}^3 \widetilde{{\cal X}}_j^{\alpha\beta}e^{-i\widetilde E_j L}
\right|^2
\nonumber\\
&{\ }&\hspace*{-16mm}
= \left|e^{-i\widetilde E_1 L}
\sum_{j=1}^3 \widetilde{{\cal X}}_j^{\alpha\beta}e^{-i\Delta \widetilde E_{j1} L}
\right|^2
\nonumber\\
&{\ }&\hspace*{-16mm}
= \left|
\sum_{j=1}^3 \widetilde{{\cal X}}_j^{\alpha\beta}
\left\{1-\left(1-e^{-i\Delta \widetilde E_{j1} L}\right)\right\}
\right|^2
\nonumber\\
&{\ }&\hspace*{-16mm}
= \left|[N^\ast N^T]_{\alpha\beta}
-2i\sum_{j=2}^3
e^{-i\Delta \widetilde E_{j1} L/2}
\widetilde{{\cal X}}_j^{\alpha\beta}
\sin\left(\frac{\Delta \widetilde E_{j1} L}{2}\right)
\right|^2
\nonumber\\
&{\ }&\hspace*{-16mm}
= \left|
[\{({\bf 1}+\eta)^2\}^T]_{\alpha\beta}
+2e^{-i\Delta \widetilde E_{31} L/2-i\pi/2}
\widetilde{{\cal X}}_3^{\alpha\beta}
\sin\left(\frac{\Delta \widetilde E_{31} L}{2}\right)\right.
\nonumber\\
&{\ }&\hspace*{14mm}
\left.+2e^{-i\Delta \widetilde E_{21} L/2-i\pi/2}
\widetilde{{\cal X}}_2^{\alpha\beta}
\sin\left(\frac{\Delta \widetilde E_{21} L}{2}\right)
\right|^2
\nonumber\\
&{\ }&\hspace*{-16mm}
\simeq 4\left|
\eta_{\beta\alpha}
+e^{-i\Delta \widetilde E_{31} L/2-i\pi/2}
\widetilde{{\cal X}}_3^{\alpha\beta}
\sin\left(\frac{\Delta \widetilde E_{31} L}{2}\right)\right.
\nonumber\\
&{\ }&\hspace*{-3mm}
\left.+e^{-i\Delta \widetilde E_{21} L/2-i\pi/2}
\widetilde{{\cal X}}_2^{\alpha\beta}
\sin\left(\frac{\Delta \widetilde E_{21} L}{2}\right)
\right|^2\,.
\label{uvprobv}
\end{eqnarray}
T violation $P(\nu_\mu\to\nu_e)-P(\nu_e\to\nu_\mu)$ 
is a small quantity, and
the difference between the probability
$P(\nu_\mu\to\nu_e)$
and the modified one $\hat{P}(\nu_\mu\to\nu_e)$
comes from the factor
$(NN^\dagger)_{\alpha\alpha}(NN^\dagger)_{\beta\beta}
=[({\bf 1}+\eta)^2]_{\alpha\alpha}
[({\bf 1}+\eta)^2]_{\beta\beta}\simeq
1+2\eta_{\alpha\alpha}+2\eta_{\beta\beta}$,
which has a small deviation from 1.
Therefore, T violation of the probability 
$P(\nu_\mu\to\nu_e)-P(\nu_e\to\nu_\mu)$ 
can be approximated by
that of the modified probability
$\hat{P}(\nu_\mu\to\nu_e)-\hat{P}(\nu_e\to\nu_\mu)$.
Hence T violation is given by
\begin{eqnarray}
&{\ }&\hspace*{-4mm}
  P(\nu_\mu\to\nu_e)-P(\nu_e\to\nu_\mu)
  \nonumber\\
&{\ }&\hspace*{-8mm}
\simeq  \hat{P}(\nu_\mu\to\nu_e)-\hat{P}(\nu_e\to\nu_\mu)
  \nonumber\\
&{\ }&\hspace*{-8mm}
=4\left|
\eta_{\mu e}
+e^{-i\Delta \widetilde E_{31} L/2-i\pi/2}
\widetilde{{\cal X}}_3^{e\mu}
\sin\left(\frac{\Delta \widetilde E_{31} L}{2}\right)
+e^{-i\Delta \widetilde E_{21} L/2-i\pi/2}
\widetilde{{\cal X}}_2^{e\mu}
\sin\left(\frac{\Delta \widetilde E_{21} L}{2}\right)
\right|^2
\nonumber\\
&{\ }&\hspace*{-7mm}
-4\left|
\eta_{e\mu}
+e^{-i\Delta \widetilde E_{31} L/2-i\pi/2}
\widetilde{{\cal X}}_3^{\mu e}
\sin\left(\frac{\Delta \widetilde E_{31} L}{2}\right)
+e^{-i\Delta \widetilde E_{21} L/2-i\pi/2}
\widetilde{{\cal X}}_2^{\mu e}
\sin\left(\frac{\Delta \widetilde E_{21} L}{2}\right)
\right|^2
\nonumber\\
&{\ }&\hspace*{-8mm}
=4\sin\left(\frac{\Delta \widetilde E_{31} L}{2}\right)
\left(\eta_{\mu e}\widetilde{{\cal X}}_3^{e\mu\ast}
-\eta_{\mu e}^\ast\widetilde{{\cal X}}_3^{e\mu}\right)
(e^{i\Delta \widetilde E_{31} L/2+i\pi/2}
-e^{-i\Delta \widetilde E_{31} L/2-i\pi/2})
\nonumber\\
&{\ }&\hspace*{-6mm}
+4\sin\left(\frac{\Delta \widetilde E_{21} L}{2}\right)
\left(\eta_{\mu e}\widetilde{{\cal X}}_2^{e\mu\ast}
-\eta_{\mu e}^\ast\widetilde{{\cal X}}_2^{e\mu}\right)
(e^{i\Delta \widetilde E_{21} L/2+i\pi/2}
-e^{-i\Delta \widetilde E_{21} L/2-i\pi/2})
\nonumber\\
&{\ }&\hspace*{-6mm}
-4\sin\left(\frac{\Delta \widetilde E_{31} L}{2}\right)
\sin\left(\frac{\Delta \widetilde E_{21} L}{2}\right)
\left(\widetilde{{\cal X}}_3^{e\mu}\widetilde{{\cal X}}_2^{e\mu\ast}
-\widetilde{{\cal X}}_3^{e\mu\ast}\widetilde{{\cal X}}_2^{e\mu}\right)
\nonumber\\
&{\ }&\hspace*{-4mm}
\times (e^{i\Delta \widetilde E_{31} L/2+i\pi/2-i\Delta \widetilde E_{21} L/2-i\pi/2}
-e^{-i\Delta \widetilde E_{31} L/2-i\pi/2+i\Delta \widetilde E_{21} L/2+i\pi/2})
\nonumber\\
&{\ }&\hspace*{-9mm}
=
-16\,\mbox{\rm Im}\left[\widetilde{{\cal X}}_2^{e\mu}\widetilde{{\cal X}}_3^{e\mu\ast}\right]
\sin\left(\frac{\Delta \widetilde E_{31} L}{2}\right)
\sin\left(\frac{\Delta \widetilde E_{21} L}{2}\right)
\sin\left(\frac{\Delta\widetilde E_{32} L}{2}\right)
\nonumber\\
&{\ }&\hspace*{-5mm}
+8\,\mbox{\rm Im}\left[\eta_{\mu e}\widetilde{{\cal X}}_3^{e\mu\ast}\right]
\sin\left(\Delta \widetilde E_{31} L\right)
+8\,\mbox{\rm Im}\left[\eta_{\mu e}\widetilde{{\cal X}}_2^{e\mu\ast}\right]
\sin\left(\Delta\widetilde E_{21} L\right)
\label{uvtv1}
\end{eqnarray}
We observe that the energy dependence of T violation
in this case is different from that with unitarity,
since we have extra contributions which are
proportional to $\sin\left(\Delta \widetilde E_{31} L\right)$
or $\sin\left(\Delta\widetilde E_{21} L\right)$.
As in the case with unitarity,
$\widetilde{{\cal X}}^{\alpha\beta}_j$ can be expressed
in terms of the quantity
$X^{\alpha\beta}_j\equiv
U_{\alpha j} U_{\beta j}^\ast$ in vacuum, $\widetilde{E}_j$,
and $\eta_{\alpha\beta}$.
First of all, we note the following relations:
\begin{eqnarray}
  &{\ }&\hspace*{-26mm}
  \sum_j(\widetilde{E}_j)^m\widetilde{{\cal X}}^{\alpha\beta}_j
\nonumber\\
&{\ }&\hspace*{-29mm}
=\sum_j(N^\ast W)_{\alpha j}
(\widetilde{E}_j)^m(NW^\ast)_{\beta j}
\nonumber\\
&{\ }&\hspace*{-29mm}
=\left[
  N^\ast \left\{{\cal E}+N^T{\cal A}N^\ast
-A_n\left(N^TN^\ast-{\bf 1}\right)
  \right\}^m N^T\right]_{\alpha\beta}
\nonumber\\
&{\ }&\hspace*{-29mm}
=\left[
  N \left\{{\cal E}+N^\dagger{\cal A}N
-A_n\left(N^\dagger N-{\bf 1}\right)
  \right\}^m N^\dagger\right]_{\beta\alpha}
\nonumber\\
&{\ }&\hspace*{-29mm}
\equiv {\cal Y}^{\alpha\beta}_{m+1}
\hspace*{19mm}
\mbox{\rm for}~m=0,1,2.
\label{simultaneous0}
\end{eqnarray}
Then we rewrite Eqs.~(\ref{simultaneous0}) as
\begin{eqnarray}
&{\ }&\hspace*{-46mm}
\sum_{m=1}^3
V_{jm}\,\widetilde{{\cal X}}^{\alpha\beta}_m
={\cal Y}^{\alpha\beta}_j\qquad\mbox{\rm for}~j=1,2,3,
\label{simultaneous1}
\end{eqnarray}
where
$V_{jm}\equiv(\widetilde{E}_m)^{j-1}$
is the element of 
the Vandermonde matrix $V$ as in the case with unitarity
(See Eq. (\ref{verdermonde})).
The simultaneous equation~(\ref{simultaneous1}) can be solved by inverting $V$
and we obtain
\begin{eqnarray}
&{\ }&\hspace*{-61mm}
\widetilde{{\cal X}}^{\alpha\beta}_j=\sum_{m=1}^3
(V^{-1})_{jm}\,{\cal Y}^{\alpha\beta}_m\,.
\label{simultaneous3}
\end{eqnarray}
The factor
$\,\mbox{\rm Im}\left[\widetilde{{\cal X}}_2^{e\mu}\widetilde{{\cal X}}_3^{e\mu\ast}\right]$
can be expressed in terms of $\widetilde E_j$ and
${\cal Y}^{e\mu}_j$:
\begin{eqnarray}
&{\ }&\hspace*{-15mm}
\mbox{\rm Im}\,\left[\widetilde{{\cal X}}_2^{e\mu}
  \widetilde{{\cal X}}_3^{e\mu\ast}\right]
\nonumber\\
&{\ }&\hspace*{-10mm}
=\displaystyle
\frac{-1}{\Delta \widetilde{E}_{21} \Delta \widetilde{E}_{32}}
\frac{{\ }1}{\Delta \widetilde{E}_{31} \Delta \widetilde{E}_{32}}
\nonumber\\
&{\ }&\hspace*{-8mm}
\times\,\mbox{\rm Im}\,\left[
  \{{\cal Y}_3^{e\mu}-(\widetilde{E}_3+\widetilde{E}_1){\cal Y}_2^{e\mu}
  +\widetilde{E}_3\widetilde{E}_1{\cal Y}_1^{e\mu}\}
\{{\cal Y}_3^{e\mu\ast}-(\widetilde{E}_1+\widetilde{E}_2)
    {\cal Y}_2^{e\mu\ast}
    +\widetilde{E}_1\widetilde{E}_2{\cal Y}_1^{e\mu\ast}
    \}\right]
\nonumber\\
&{\ }&\hspace*{-10mm}
=\displaystyle
\frac{1}{\Delta \widetilde{E}_{21} \Delta \widetilde{E}_{31}
  \Delta \widetilde{E}_{32}}
\left(\widetilde{E}_1^2\,\mbox{\rm Im}\left[{\cal Y}^{e\mu}_1{\cal Y}^{e\mu\ast}_2\right]
-\widetilde{E}_1 \,\mbox{\rm Im}\left[{\cal Y}^{e\mu}_1{\cal Y}^{e\mu\ast}_3\right]
+\,\mbox{\rm Im}\left[{\cal Y}^{e\mu}_2{\cal Y}^{e\mu\ast}_3\right]\right)\,.
\label{uvtv2}
\end{eqnarray}
The quantities ${\cal Y}^{e\mu}_j~(j=1,2,3)$ are calculated as follows:
\begin{eqnarray}
  &{\ }&\hspace*{-76mm}
  {\cal Y}^{e\mu}_1 =
  [N N^\dagger]_{\mu e}
  =[({\bf 1}+\eta)^2]_{\mu e}\simeq 2\,\eta_{\mu e}\,,
\nonumber
\end{eqnarray}
\begin{eqnarray}
&{\ }&\hspace*{-16mm}
  {\cal Y}^{e\mu}_2 =\left[
    N\left\{{\cal E}+N^\dagger{\cal A}\,N
-A_n\left(N^\dagger N-{\bf 1}\right)
    \right\} N^\dagger\right]_{\mu e}
\nonumber\\
&{\ }&\hspace*{-8mm}
=\left.
\left[({\bf 1}+\eta)
  \left\{U{\cal E}U^{-1}+({\bf 1}+\eta){\cal A}({\bf 1}+\eta)
-A_n\left(({\bf 1}+\eta)^2-{\bf 1}\right)
  \right\}
  ({\bf 1}+\eta)\right]\right|_{\mu e}
\nonumber\\
&{\ }&\hspace*{-8mm}
\simeq\left[U{\cal E}U^{-1}+{\cal A}+\{\eta,U{\cal E}U^{-1}\}
  +2\{\eta,{\cal A}\}-2A_n\eta\right]_{\mu e}
\nonumber\\
&{\ }&\hspace*{-8mm}
\simeq\Delta E_{31} X_3^{\mu e}
+\Delta E_{21} X_2^{\mu e}
+\Delta E_{31} \{\eta,X_3\}_{\mu e}
+2(A-A_n)\eta_{\mu e}\,,
\nonumber
\end{eqnarray}
\begin{eqnarray}
&{\ }&\hspace*{-15mm}
{\cal Y}^{e\mu}_3 =
\left[
    N\left\{{\cal E}+N^\dagger{\cal A}\,N
-A_n\left(N^\dagger N-{\bf 1}\right)
    \right\}^2 N^\dagger\right]_{\mu e}
\nonumber\\
&{\ }&\hspace*{-8mm}
=\left.
\left[({\bf 1}+\eta)
  \left\{U{\cal E}U^{-1}+({\bf 1}+\eta){\cal A}({\bf 1}+\eta)
-A_n\left(({\bf 1}+\eta)^2-{\bf 1}\right)
  \right\}^2
  ({\bf 1}+\eta)\right]\right|_{\mu e}
\nonumber\\
&{\ }&\hspace*{-8mm}
\simeq\left[U{\cal E}^2U^{-1}+{\cal A}^2
  +\{U{\cal E}U^{-1},{\cal A}\}\right]_{\mu e}
  +\left[\{U{\cal E}U^{-1},\{\eta,{\cal A}\}\}
+\{{\cal A},\{\eta,{\cal A}\}\}\right]_{\mu e}
\nonumber\\
&{\ }&\hspace*{-6mm}
  -\left[2A_n\{U{\cal E}U^{-1},\eta\}
+2A_n\{{\cal A},\eta\}\right]_{\mu e}
\nonumber\\
&{\ }&\hspace*{-6mm}
+\left[\{\eta,U{\cal E}^2 U^{-1}\}+\{\eta,{\cal A}^2\}
  +\{\eta,\{U{\cal E}U^{-1},{\cal A}\}\}\right]_{\mu e}
\nonumber\\
&{\ }&\hspace*{-8mm}
\simeq\Delta E_{31}(\Delta E_{31}+A)\,X_3^{\mu e}
+A\,\Delta E_{21}X_2^{\mu e}
\nonumber\\
&{\ }&\hspace*{-3mm}
+\Delta E_{31}\{X_3,\{\eta,{\cal A}\}\}_{\mu e}
+ \{{\cal A},\{\eta,{\cal A}\}\}_{\mu e}
\nonumber\\
&{\ }&\hspace*{-3mm}
  -2A_n\Delta E_{31}\{X_3,\eta\}_{\mu e}
-2A_n\{{\cal A},\eta\}_{\mu e}
\nonumber\\
&{\ }&\hspace*{-3mm}
+ \Delta E_{31}^2\{\eta,X_3\}_{\mu e}
  +\{\eta,{\cal A}^2\}_{\mu e}
  +\Delta E_{31}\{\eta,\{X_3,{\cal A}\}\}_{\mu e}
\nonumber\\
&{\ }&\hspace*{-8mm}
=(\Delta E_{31}+A)\,{\cal Y}^{e\mu}_2
-\Delta E_{31}\Delta E_{21}\, X_2^{\mu e}
\nonumber\\
&{\ }&\hspace*{-3mm}
-2(A-A_n)\Delta E_{31}\eta_{\mu e}
-(A+2A_n)\Delta E_{31}\{X_3,\eta\}_{\mu e}
\nonumber\\
&{\ }&\hspace*{-3mm}
+\Delta E_{31}\{X_3,\{\eta,{\cal A}\}\}_{\mu e}
+\,\Delta E_{31}\{\eta,\{X_3,{\cal A}\}\}_{\mu e}\,.
\nonumber
\end{eqnarray}
In the current scenario involving unitarity violation, we observe a
nonvanishing contribution from ${\cal Y}^{\alpha\beta}_1$,
necessitating knowledge of the explicit form of the energy eigenvalue
$\widetilde{E}_1$.  Given that ${\cal Y}^{e\mu}_1$ is of order
$O(\eta_{\alpha\beta})$, to evaluate Eq. (\ref{uvtv2}) accurately to
first order in both $\Delta E_{21}/\Delta E_{31}$ and
$\eta_{\alpha\beta}$, we must calculate $\widetilde{E}_1$ solely to
zeroth order in these parameters, i.e., assuming $\Delta E_{21}\to 0$
and $\eta_{\alpha\beta}\to 0$.  Under these conditions, the
characteristic equation of the $3\times 3$ matrix (\ref{uvm}) is
defined by
\begin{eqnarray}
  &{\ }&\hspace*{-36mm}
0=\mbox{\rm det}\left[{\bf 1}\,t-
  \left\{{\cal E}+N^T{\cal A}N^\ast
  -A_n\left(N^TN^\ast-{\bf 1}\right)
\right\}\right]
\nonumber\\
&{\ }&\hspace*{-33mm}
\simeq\mbox{\rm det}\left[{\bf 1}\,t
-\mbox{\rm diag}(0,0,\Delta E_{31})
-U^{-1}\,\mbox{\rm diag}(A,0,0)\,U\right]
\nonumber\\
&{\ }&\hspace*{-33mm}
=t\,\left\{t^2-(\Delta E_{31} + A)\,t
+ A\Delta E_{31}\cos^2\theta_{13}\right\}
\nonumber\\
&{\ }&\hspace*{-33mm}
=t\,(t-\lambda_+)(t-\lambda_-)\,,
\nonumber
\end{eqnarray}
where $\lambda_\pm$ are the roots of the quadratic equation
and are given by
\begin{eqnarray}
&{\ }&\hspace*{-33mm}
  \lambda_\pm\equiv
  \frac{\Delta E_{31} + A\pm\Delta \widetilde{E}_{31}}{2}
\nonumber\\
&{\ }&\hspace*{-33mm}
\Delta \widetilde{E}_{31}\equiv
\sqrt{(\Delta E_{31}\cos2\theta_{13}-A)^2+(\Delta E_{31}\sin2\theta_{13})^2}\,.
\nonumber
\end{eqnarray}
From this, we obtain the energy eigenvalues
$\widetilde{E}_j~(j=1,2,3)$ to the leading order in
$\Delta E_{21}/\Delta E_{31}$ and $\eta_{\alpha\beta}$:
\begin{eqnarray}
  &{\ }&\hspace*{-86mm}
\left(
\begin{array}{c}
  \widetilde{E}_1\\
  \widetilde{E}_2\\
  \widetilde{E}_3
\end{array}
\right)
\simeq\left(
\begin{array}{c}
  \lambda_-\\
  0\\
  \lambda_+
  \end{array}
\right)
\nonumber
\end{eqnarray}
The roots $\lambda_-=\widetilde{E}_1$ and
$\lambda_+=\widetilde{E}_3$ satisfy
of the quadratic equation
\begin{eqnarray}
  &{\ }&\hspace*{-46mm}
  \lambda_\pm^2 - (\Delta E_{31}+A)\lambda_\pm
  + A\Delta E_{31} \cos^2\theta_{13} = 0\,.
\nonumber
\end{eqnarray}
Hence the first two term on the right hand
side of Eq. (\ref{uvtv2}) can be rewritten as
\begin{eqnarray}
&{\ }&\hspace*{-46mm}
\widetilde{E}_1^2\,\mbox{\rm Im}\left[{\cal Y}^{e\mu}_1{\cal Y}^{e\mu\ast}_2\right]
-\widetilde{E}_1 \,\mbox{\rm Im}\left[{\cal Y}^{e\mu}_1{\cal Y}^{e\mu\ast}_3\right]
\nonumber\\
&{\ }&\hspace*{-51mm}
=\mbox{\rm Im}\left[{\cal Y}^{e\mu}_1\left\{
  \widetilde{E}_1^2{\cal Y}^{e\mu\ast}_2
  -\widetilde{E}_1{\cal Y}^{e\mu\ast}_3
  \right\}\right]
\nonumber\\
&{\ }&\hspace*{-51mm}
\simeq\,\mbox{\rm Im}\left[{\cal Y}^{e\mu}_1\left\{
  \Delta E_{31} X_3^{\mu e\ast}\widetilde{E}_1^2
  -\Delta E_{31}(\Delta E_{31}+A)\, X_3^{\mu e\ast}\widetilde{E}_1
  \right\}\right]
\nonumber\\
&{\ }&\hspace*{-51mm}
=\,\mbox{\rm Im}\left[{\cal Y}^{e\mu}_1
  A(\Delta E_{31})^2X_3^{\mu e\ast}\cos^2\theta_{13} \right]
\nonumber\\
&{\ }&\hspace*{-51mm}
=2\,A\,(\Delta E_{31})^2\,\cos^2\theta_{13}\,\mbox{\rm Im}
\left[\eta_{\mu e}X_3^{\mu e\ast}\right]\,,
\nonumber
\end{eqnarray}
whereas the third term on the right hand
side of Eq. (\ref{uvtv2}) can be written as
\begin{eqnarray}
&{\ }&\hspace*{-25mm}
  \mbox{\rm Im}\left[{\cal Y}^{e\mu}_2{\cal Y}^{e\mu\ast}_3\right]
\nonumber\\
&{\ }&\hspace*{-29mm}
=-\,\mbox{\rm Im}\left[{\cal Y}^{e\mu\ast}_2
  \left\{
(\Delta E_{31}+A)\,{\cal Y}^{e\mu}_2
-\Delta E_{31}\Delta E_{21}\, X_2^{\mu e}
\right.\right.
\nonumber\\
&{\ }&\hspace*{-17mm}
-2(A-A_n)\Delta E_{31}\eta_{\mu e}
-(A+2A_n)\Delta E_{31}\{X_3,\eta\}_{\mu e}
\nonumber\\
&{\ }&\hspace*{-17mm}
\left.\left.
+\Delta E_{31}\{X_3,\{\eta,{\cal A}\}\}_{\mu e}
+\,\Delta E_{31}\{\eta,\{X_3,{\cal A}\}\}_{\mu e}
  \right\}
  \right]
\nonumber\\
&{\ }&\hspace*{-29mm}
=\Delta E_{31}\,\mbox{\rm Im}\left[X_3^{\mu e\ast}
  \left\{
\Delta E_{31}\Delta E_{21} X_2^{\mu e}
\right.\right.
\nonumber\\
&{\ }&\hspace*{-1mm}
+2(A-A_n)\Delta E_{31}\eta_{\mu e}
+(A+2A_n)\Delta E_{31}\{X_3,\eta\}_{\mu e}
\nonumber\\
&{\ }&\hspace*{-1mm}
\left.\left.
 -\Delta E_{31}\{X_3,\{\eta,{\cal A}\}\}_{\mu e}
 -\Delta E_{31}\{\eta,\{X_3,{\cal A}\}\}_{\mu e}
  \right\}
  \right]\,.
\nonumber
\end{eqnarray}
Thus we obtain the expression for the factor
$\,\mbox{\rm Im}\left[\widetilde{{\cal X}}_2^{e\mu}
\widetilde{{\cal X}}_3^{e\mu\ast}\right]$:
\begin{eqnarray}
&{\ }&\hspace*{-25mm}
\mbox{\rm Im}\,\left[\widetilde{{\cal X}}_2^{e\mu}
  \widetilde{{\cal X}}_3^{e\mu\ast}\right]
\nonumber\\
&{\ }&\hspace*{-30mm}
  =\displaystyle
\frac{1}{\Delta \widetilde{E}_{21} \Delta \widetilde{E}_{31}
  \Delta \widetilde{E}_{32}}
\left(\widetilde{E}_1^2\,\mbox{\rm Im}\left[{\cal Y}^{e\mu}_1{\cal Y}^{e\mu\ast}_2\right]
-\widetilde{E}_1 \,\mbox{\rm Im}\left[{\cal Y}^{e\mu}_1{\cal Y}^{e\mu\ast}_3\right]
+\,\mbox{\rm Im}\left[{\cal Y}^{e\mu}_2{\cal Y}^{e\mu\ast}_3\right]\right)
\nonumber\\
&{\ }&\hspace*{-30mm}
\simeq
\frac{\Delta E_{31}}{\Delta \widetilde{E}_{21} \Delta \widetilde{E}_{31}
  \Delta \widetilde{E}_{32}}
\,\mbox{\rm Im}\left[X_3^{\mu e\ast}
  \left\{
\Delta E_{21} X_2^{\mu e}
+4\,A\,\eta_{\mu e}
+2A_n \left(\{\eta,X_3\}_{\mu e}
-\eta_{\mu e}\right)
\right\}\right]
\label{uvtv3}
\end{eqnarray}
To complete the calculation of Eq. (\ref{uvtv1}),
we need to estimate the two quantities:
\begin{eqnarray}
&{\ }&\hspace*{-35mm}
\mbox{\rm Im}\left[\eta_{\mu e}\widetilde{{\cal X}}_3^{e\mu\ast}\right]
\nonumber\\
&{\ }&\hspace*{-40mm}
=\frac{1}{\Delta \widetilde{E}_{31} \Delta \widetilde{E}_{32}}
\,\mbox{\rm Im}\left[\eta_{\mu e}\left\{
  \widetilde{E}_1\widetilde{E}_2{\cal Y}^{e\mu\ast}_1
  -(\widetilde{E}_1+\widetilde{E}_2){\cal Y}^{e\mu\ast}_2
  +{\cal Y}^{e\mu\ast}_3
  \right\}\right]
\nonumber\\
&{\ }&\hspace*{-40mm}
\simeq\frac{1}{\Delta \widetilde{E}_{31} \Delta \widetilde{E}_{32}}
\,\mbox{\rm Im}\left[\eta_{\mu e}\left(
  -\widetilde{E}_1{\cal Y}^{e\mu\ast}_2
  +{\cal Y}^{e\mu\ast}_3
  \right)\right]
\nonumber\\
&{\ }&\hspace*{-40mm}
\simeq\frac{1}{\Delta \widetilde{E}_{31} \lambda_+}
\,\mbox{\rm Im}\left[\eta_{\mu e}\left(
  -\lambda_-\Delta E_{31} X_3^{\mu e\ast}
  +\Delta E_{31}(\Delta E_{31}+A) X_3^{\mu e\ast}
  \right)\right]
\nonumber\\
&{\ }&\hspace*{-40mm}
=\frac{1}{\Delta \widetilde{E}_{31} \lambda_+}
\lambda_+\,\Delta E_{31}\,\mbox{\rm Im}\left[\eta_{\mu e}X_3^{\mu e\ast}\right]
\nonumber\\
&{\ }&\hspace*{-40mm}
=\frac{\Delta E_{31}}{\Delta \widetilde{E}_{31}}
\,\mbox{\rm Im}\left[\eta_{\mu e}X_3^{\mu e\ast}\right]\,,
\label{uvtv4}
\end{eqnarray}
\begin{eqnarray}
&{\ }&\hspace*{-35mm}
\mbox{\rm Im}\left[\eta_{\mu e}\widetilde{{\cal X}}_2^{e\mu\ast}\right]
\nonumber\\
&{\ }&\hspace*{-40mm}
=\frac{-1}{\Delta \widetilde{E}_{21} \Delta \widetilde{E}_{32}}
\,\mbox{\rm Im}\left[\eta_{\mu e}\left\{
  \widetilde{E}_3\widetilde{E}_1{\cal Y}^{e\mu\ast}_1
  -(\widetilde{E}_3+\widetilde{E}_1){\cal Y}^{e\mu\ast}_2
  +{\cal Y}^{e\mu\ast}_3
  \right\}\right]
\nonumber\\
&{\ }&\hspace*{-40mm}
\simeq\frac{-1}{\Delta \widetilde{E}_{21} \Delta \widetilde{E}_{32}}
\,\mbox{\rm Im}\left[-(\widetilde{E}_3+\widetilde{E}_1){\cal Y}^{e\mu\ast}_2
  +{\cal Y}^{e\mu\ast}_3
\right]
\nonumber\\
&{\ }&\hspace*{-40mm}
\simeq\frac{-1}{\Delta \widetilde{E}_{21} \Delta \widetilde{E}_{32}}
\,\mbox{\rm Im}\left[-(\widetilde{E}_3+\widetilde{E}_1)
  \Delta E_{31} X_3^{\mu e\ast}
  +\Delta E_{31}(\Delta E_{31}+A) X_3^{\mu e\ast}\right]
\nonumber\\
&{\ }&\hspace*{-40mm}
\simeq 0\,,
\label{uvtv5}
\end{eqnarray}
where terms of order
$O((\Delta E_{21}/\Delta E_{31})^2)$,
$O((\epsilon_{\alpha\beta})^2)$
and $O(\epsilon_{\alpha\beta}\Delta E_{21}/\Delta E_{31})$
have been neglected in Eqs. (\ref{uvtv3}), (\ref{uvtv4}) and (\ref{uvtv5}).
Putting Eqs. (\ref{uvtv3}), (\ref{uvtv4}) and (\ref{uvtv5})
together, the final expression for T violation is given by
\begin{eqnarray}
&{\ }&\hspace*{-15mm}
  P(\nu_\mu\to\nu_e)-P(\nu_e\to\nu_\mu)
  \nonumber\\
&{\ }&\hspace*{-19mm}
\simeq
16\frac{\Delta E_{31}}{A\,\Delta \widetilde{E}_{31}\,\cos^2\theta_{13}}
\sin\left(\frac{\Delta \widetilde E_{31} L}{2}\right)
\sin\left(\frac{\Delta \widetilde E_{21} L}{2}\right)
\sin\left(\frac{\Delta\widetilde E_{32} L}{2}\right)
\nonumber\\
&{\ }&\hspace*{-14mm}
\times \,\mbox{\rm Im}\left[X_3^{\mu e\ast}
  \left\{
\Delta E_{21} X_2^{\mu e}
+4\,A\,\eta_{\mu e}
+2A_n \left(\{\eta,X_3\}_{\mu e}
-\eta_{\mu e}\right)
\right\}\right]
\nonumber\\
&{\ }&\hspace*{-17mm}
+8\,\frac{\Delta E_{31}}{\Delta \widetilde{E}_{31}}
\sin\left(\Delta \widetilde E_{31} L\right)
\,\mbox{\rm Im}\left[\eta_{\mu e}X_3^{\mu e\ast}\right]
\label{uvtv6}
\end{eqnarray}
Due to the additional contribution proportional to $\sin(\Delta
\widetilde E_{31} L)$, the energy dependence of Eq. (\ref{uvtv6}) in
the scenario with unitarity violation differs from that in the
scenarios with unitarity, such as the standard case (\ref{stdtv2}) and
the nonstandard interaction case (\ref{uvtv6}).  Therefore, if the
contribution from unitarity violation is significant enough and the
experimental sensitivity is sufficiently high, it may be possible to
distinguish the unitarity violation scenario from both the standard
and nonstandard interaction scenarios by examining the energy spectrum
in T violation.

\section{Conclusions \label{conclusions}}
In this paper, we have derived the analytical expression for T
violation in neutrino oscillations under three different scenarios:
the standard three flavor mixing framework, a scenario involving
flavor-dependent nonstandard interactions, and a case with unitarity
violation.  In scenarios preserving unitarity, the T-violating
component of the oscillation probability is proportional to
$\sin(\frac{\Delta \widetilde E_{31} L}{2}) \sin(\frac{\Delta
\widetilde E_{21} L}{2}) \sin(\frac{\Delta\widetilde E_{32} L}{2})$.
However, in the case with unitarity violation, there is an additional
contribution proportional to $\sin(\Delta \widetilde E_{31} L)$.
Should future long-baseline experiments{, such as $\mu$TRISTAN
or other types of neutrino factories,} achieve high sensitivity to T
violation across a broad energy spectrum, it may become feasible to
specifically probe unitarity in the $\nu_\mu\leftrightarrow\nu_e$
channel.

Moreover, we demonstrated that the coefficient of the term\\
$\sin(\frac{\Delta \widetilde E_{31} L}{2}) \sin(\frac{\Delta
\widetilde E_{21} L}{2}) \sin(\frac{\Delta\widetilde E_{32} L}{2})
(\Delta E_{31})^2 (\Delta \widetilde E_{31} \Delta \widetilde E_{32}
\Delta \widetilde E_{21})^{-1}$ varies depending on whether neutrino
propagation follows the standard scheme or involves nonstandard
interactions.  In the standard scenario, this coefficient is
proportional to $\Delta E_{21} = \Delta m_{21}^2/2E$.  However, in the
case with nonstandard interactions, there is an additional
contribution that is energy-independent.  Thus, it may be possible to
observe the effects of nonstandard interactions by examining the
energy dependence of T violation.

The purpose of this paper is to derive the analytical expression of T
violation, and we did not quantitatively discuss the sensitivity of
future experiments.  The potential for T violation in neutrino
oscillations deserves further study.
  
\section*{Acknowledgments}
From 1989 to 1991, I was a postdoc at the University of North
Carolina at Chapel Hill, and I would like to thank Prof. Frampton
for giving me the opportunity to conduct research at UNC. I am
delighted to celebrate Prof. Frampton's 80th birthday and wish
him continued success and activity in the years to come.
This research was partly supported by a
Grant-in-Aid for Scientific Research of the Ministry of Education,
Science and Culture, under Grant No. 21K03578.

\end{document}